\documentclass[aps,prb,twocolumn,groupedaddress]{revtex4}

\usepackage{graphicx}

\newcommand{\bs}{\begin{scriptsize}}
\newcommand{\es}{\end{scriptsize}}

\begin{document}


\title{A simple and controlled single electron transistor \linebreak based on doping modulation in silicon nanowires}


\author{M. Hofheinz, X. Jehl and M. Sanquer}
\affiliation{DSM-DRFMC-SPSMS, CEA-Grenoble}
\email[]{xavier.jehl@cea.fr}
\author{G. Molas, M. Vinet and S. Deleonibus}
\affiliation{DRT-LETI-D2NT-LNDE, CEA-Grenoble}


\begin{abstract}

A simple and highly reproducible single electron transistor (SET) has been fabricated using gated silicon nanowires. The structure is a metal-oxide-semiconductor field-effect transistor made on silicon-on-insulator thin films. The channel of the transistor is the Coulomb island at low temperature. Two silicon nitride spacers deposited on each side of the gate create a modulation of doping along the nanowire that creates tunnel barriers. Such barriers are fixed and controlled, like in metallic SETs. The period of the Coulomb oscillations is set by the gate capacitance of the transistor and therefore controlled by lithography. The source and drain capacitances have also been characterized. This design could be used to build more complex SET devices.

\end{abstract}


\maketitle


The first and most common Coulomb blockade device is the Single Electron Transistor (SET) made with metallic leads and island, and tunnel oxide barriers \cite{fulton,livre92}. It is used as sensitive electrometers \cite{devoret} or electron pumps allowing to control the transfer of electrons one by one \cite{pothier,nist}. Since then very important efforts have been devoted to fabricate silicon SETs, mostly to integrate SETs together with regular transistors for building logic circuits \cite{hadley,ono}, and more recently for quantum logic experiments with single charge or spin in silicon quantum dots \cite{gorman,aussies}. An important challenge is to increase the temperature of operation from the typical sub-kelvin range of original devices up to much higher temperatures. The required size of the island is of the order of the nm, and therefore out of control of current fabrication processes. Researchers took advantage of natural disorder to create such extremely small islands, mostly with constrictions in disordered thin films \cite{Ish97,Sak98,Aug00,Wan02,Tan03}. More recently\cite{Uch03} undulated thin films have been used, as well as pattern-dependent oxidation \cite{nishiguchi}.

\begin{figure}
\includegraphics[width=\columnwidth]{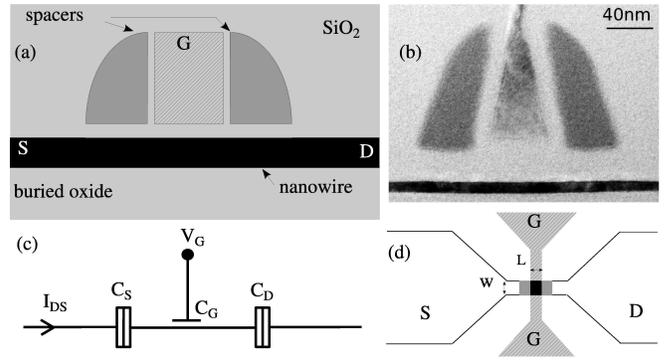}
\caption{(a) Schematics of our transistor (cross section along the drain-source axis) with the gate (G), source (S) and drain (D). (b) Transmission electron microscope image of a device, corresponding to (a). (c) Electrostatic equivalent circuit. The tunnel barriers under the spacers replace the tunnel oxide junctions (boxes) of metallic SETs, and the transistor gate replaces the usual electrostatic capacitor. (d) Top view of the device showing the nanowire that is heavily doped except under the spacers (grey areas: tunnel barriers) and under the gate (black area: Coulomb island).}
\label{device}
\end{figure}

\begin{figure*}[!t]
\includegraphics[width=\textwidth]{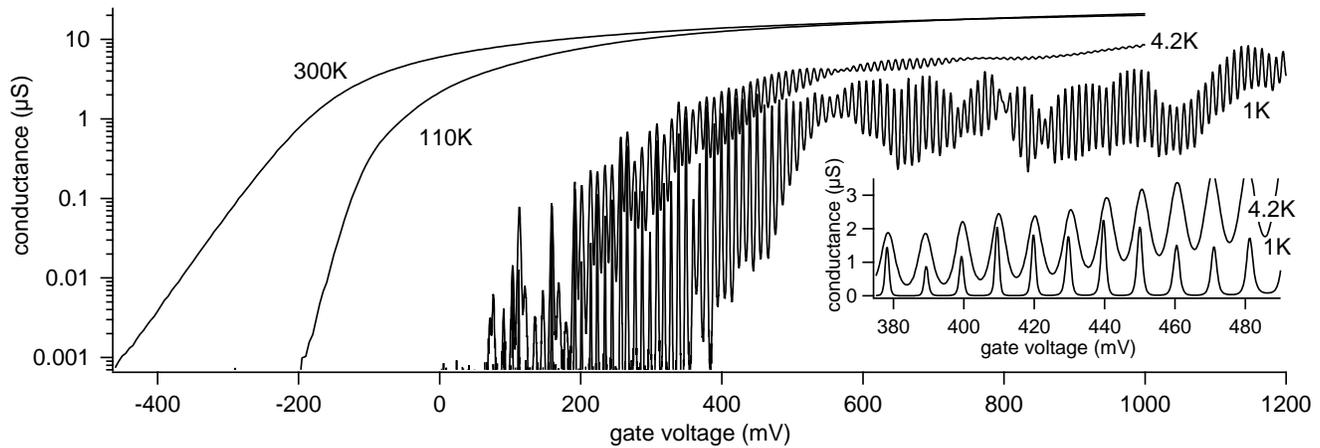}
\caption{Linear drain-source conductance versus gate voltage in a W=40\,nm, L=30\,nm device with t$_{ox}$=10\,nm, at various temperatures. Very small drain-source voltages are necessary at low temperature to stay in the linear regime: $V_{ds}$ is 500\,$\mu V$ at 300\,K, 100$\,\mu V$ at 110\,K and 4.2\,K and 80$\,\mu V$ at 1\,K. Inset: zoom on periodic Coulomb oscillations in linear scale.}
\label{allT}
\end{figure*}

We followed another approach based on etched silicon nanowires without constrictions, as pionnered by Tilke {\it et al.} \cite{tilke} and more recently Namatsu {\it et al.} \cite{Nam03}, and also Kim {\it et al.} \cite{Kim01} and Fujiwara {\it et al.} \cite{fujiwara}. In the two first cases the formation of a Coulomb island in a nanowire underneath a very large gate was studied. 
In the two others two gates were defined above a nanowire, each of them acting as a tunable barrier for entering/exiting the single electron box delimited by these gates. Although this scheme allowed logic operations to be performed at 300K \cite{Uch03,nishiguchi}, it remains a complex architecture since up to 4 gates are needed for proper operation.
Our SET is much simpler since it requires a single gate to define the quantum dot, while the barriers are fixed, like in metallic SETs. Periodic Coulomb blockade is observed, with a period solely determined by the surface area of a single gate. It is therefore controlled by lithography, not by disorder. With current state of the art electron beam lithography the limit in operating temperature is of the order of 10\,K. The schematics of our device is essentially similar to the original metallic SETs, the tunnel oxide barriers being replaced by low-doped regions under spacers producing a doping modulation along the wire, while the electrostatic gate capacitor is replaced by the MOS gate of the transistor (see Fig.~\ref{device}). We first used single electron charging effects in silicon nanocristals embedded within a gate oxide stack to fabricate few-electron memories \cite{molas}. Interestingly the use of doping modulation along a nanowire has already been used in grown devices \cite{bottom-up}, where the control of dopants is easier than ion implantation in very thin films.

\begin{figure}
\includegraphics[width=\columnwidth,clip]{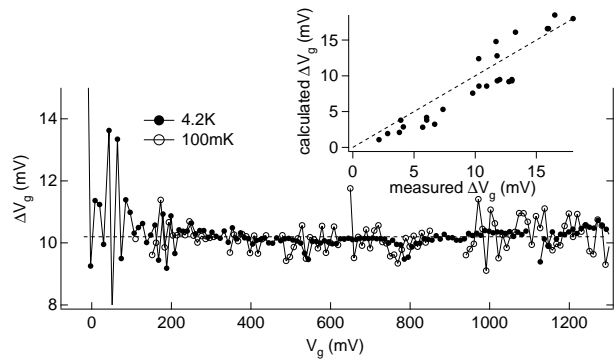}
\caption{Period of the oscillations in gate voltage $\Delta V_{g}$ in a device with W=30\,nm, L=40\,nm and t$_{ox}$=10\,nm, measured down to T=100\,mK with a drain-source voltage of 30\,$\mu V$. The gate capacitance is independent on gate voltage. 115 peaks are recorded in the gate voltage range 107 to 1293\,mV, yielding a mean spacing $\left\langle \Delta V_{g}\right\rangle $=10.2\,mV and $C_g$=15.7\,aF. Inset: calculated period $\Delta V_{g}=e/C_{g}$ from the geometry of 26 devices (with different widths, lengths and gate oxide thickness) compared to measurements. The good agreement shows that the period is set by the MOS gate capacitance.}
\label{oscill}
\end{figure}

The design is a silicon on insulator (SOI) thin film transistor fabricated on a 200\,mm CMOS platform.
First a SOI film is thinned down to approximately 17\,nm, then 200\,nm long nanowires are defined by e-beam lithography and wet etching (see Fig.~\ref{device}). The width is as small as 30\,nm for the presented devices. A first low doping of the whole nanowire and access is performed at this stage. A SiO$_2$/Poly-Si gate stack is then defined perpendicularly to the wire, with a length as small as 30\,nm. Doping modulation is achieved by using the gate as a mask for subsequent ion implantation. In previous designs we only used this gate and doped the nanowire moderatly. In a second one we added 50\,nm long silicon nitride spacers on both side of the gate and heavily doped all the uncovered regions. This scheme gives much more regular oscillations and lower background charge noise and is the one presented here.
The high doping of the uncovered wire creates low resistance contacts, and the MOSFET gate allows to accumulate electrons in the channel created under the gate. In between these regions (i.e. under the spacers) are the low doped 'access region' acting as tunnel barriers on both sides of the channel (see Fig.~\ref{device}c). We already observed Coulomb blockade in non-overlapped MOSFETs \cite{boeuf}.
The $I_{d}-V_{g}$ characteristics are shown in Fig.~\ref{allT} at various temperature. The FET characteristics at 300\,K is replaced at lower temperature by very periodic and perfectly reproducible oscillations, with a contrast increasing as the temperature decreases (see Fig.~\ref{oscill}). The period $\Delta V_{g}=e/C_{g}$ is an extremely sensitive, in-situ measurement of the gate capacitance: capacitances smaller than 10\,aF are easily measured, with very small signals. We checked that the Coulomb island is the channel of the FET transistor by comparing the measured period with the calculated MOS gate capacitance. The results are shown in the inset of Fig.~\ref{oscill} for samples with various geometries and three gate oxide thicknesses: $t_{\rm{ox}}$=4, 10 and 24\,nm. As expected the measured period scales with the surface area of the channel and the gate oxide. We believe the observed dispersion is mostly due to approximations in the simple calculation we used for a capacitor with metallic electrodes. More accurate estimations should include one semiconducting electrode, but nevertheless estimating the spatial extent of the quantum dot below the spacers remains a challenge.

\begin{figure}[!t]
\includegraphics[width=\columnwidth,clip]{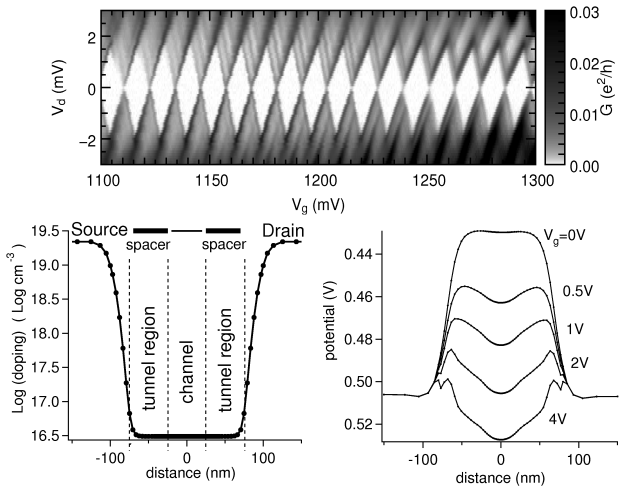}
\caption{Top: Coulomb diamonds in a device with W=60\,nm, L=40\,nm and t$_{ox}$=24\,nm. The slopes allow to determine the source and drain capacitances $C_{s}\approx$42\,aF and $C_{d}\approx$32\,aF. Bottom: numerical simulations of the doping (left) and potential (right) along the wire, 1\,nm deep below the Si/SiO$_2$ interface, for a 40\,nm gate. The undoped regions below spacers and gate create a flattop potential that is lowered in its center by the gate voltage, creating a well isolated by two barriers.}
\label{capas}
\end{figure}

Although it is well known theoretically and demonstrated experimentally \cite{krupenin} that Coulomb blockade occurs with resistive access, it is difficult in this case to predict the source and drain capacitances ($C_{s}$ and $C_{d}$), and hence the charging energy. Here the barriers do not rely on 1D constrictions or impurities, instead the wire is low doped, on the insulating side of the metal-insulator transition at 0K, providing enough resistance for confinement. We calculated the doping profile and potential along the wire at 300\,K (see Fig. \ref{capas}). The doping drops abruptly below the spacers and gate, inducing a flattop potential barrier at zero gate voltage. Increasing this voltage creates a well that pushes back down the potential in the center of the low doped region. This potential well is responsible for single electron effects at low temperature. If the resistance of this region is below a threshold value (typically h/e$^{2} \approx$ 26\,k$\Omega$), confinement does not occur. For this reason periodic Coulomb oscillations are not observed in very wide devices. On the other extreme very high access resistances result in vanishingly low current. The optimum range for our SET device is when the resistance is typically of the order of 100\,k$\Omega$ at large gate voltage (see Fig. \ref{allT}). Beyond the resistance we have characterized the source and drain capacitances by measurements in the non-linear regime. We observed very stable Coulomb diamond features, as shown in the upper panel of Fig. \ref{capas}. The slope of the rhombuses is a direct measurement of $C_{s}$ and $C_{d}$ as we already know the gate capacitance. For the data shown in Fig. \ref{capas} we found $C_{g}\approx$~13.6\,aF from the peak spacing and $C_{s}\approx$~42\,aF and $C_{d}\approx$~32\,aF from the Coulomb diamonds slopes. A major issue and a limitation in SET devices comes from switching background charges inducing a large 1/f noise and changing the phase of the Coulomb oscillations. We observe very low noise in our devices: once at cryogenic temperatures the phase of the oscillations is stable within measurement uncertainties, as long as large gate voltage sweeps (larger than 1 volt) are not performed. The few anomalies we observe are attributed to charge traps with sufficiently fast dynamics to not increase the noise within the bandwith of our measurement~\cite{max}.

In conclusion, we have shown that a precise control of doping along etched silicon nanowires allows to make a very simple, reliable and predictable single-gate SET. The period is entirely set by the transistor gate capacitance, i.e.\ by lithography. Furthermore we have characterized the source and drain capacitances arising from the low doped regions under gate spacers. We believe this basic device will be useful to realize more complex circuits fully compatible with CMOS technology.

This work was partially supported by the European Community through SiNANO (IST-506844).

\end{document}